\documentclass[10pt]{article} 
\usepackage{amsfonts,amssymb,slashed,makeidx,latexsym,setspace}
\usepackage{graphicx,graphics,amssymb,rotate} 
\usepackage{times,cite,color,epsfig}
\usepackage{hyperref}
\def\five{\bf 5}
\def\fivebar{\bf \bar 5}
\def\ten{\bf 10}

\def\24{\bf 24}

\begin{document}

\begin{flushright}
IPMU-18-0077, TU-1060
\end{flushright}

\begin{center}
{\Large \bf An extended gauge mediation for muon $\mathbf{(g-2)}$ explanation} \\
\vspace*{1cm} \renewcommand{\thefootnote}{\fnsymbol{footnote}} { {\sf
Gautam Bhattacharyya${}^{1}$}, {\sf Tsutomu
    T. Yanagida${}^{2}\footnote{Hamamatsu Professor}$}, and 
{\sf Norimi Yokozaki${}^{3}$}
}
\\
\vspace{10pt} {\small ${}^{1)}$ {\em Saha Institute of Nuclear
    Physics, HBNI, 1/AF Bidhan Nagar, Kolkata 700064, India}
  \\ ${}^{2)}$ {\em Kavli IPMU (WPI), UTIAS, University of Tokyo, Kashiwa
    277-8583, Japan} \\ ${}^{3)}$ {\em Department of Physics, Tohoku
    University, Sendai, Miyagi 980-8578, Japan} \\ } \normalsize
\end{center}

\begin{abstract}

\noindent
It is increasingly becoming difficult, within a broad class of
supersymmetric models, to satisfactorily explain the discrepancy
between the measured $(g-2)_\mu$ and its standard model prediction,
and at the same time satisfy all the other constraints.  In this paper
we propose a new scheme of gauge mediation by introducing new soft
supersymmetry breaking mass parameters for the Higgs sector in a 
minimal setup containing only a pair of $(\five + \fivebar)$ messenger
fields of $SU(5)$.  This enables us to explain the $(g-2)_\mu$
discrepancy while avoiding all the existing constraints.  We also
provide possible dynamical origin of the new soft mass parameters.
The wino and higgsino weighing below 500 GeV constitute the smoking gun signal at the (high luminosity) LHC.
\end{abstract}

\setcounter{footnote}{0}
\renewcommand{\thefootnote}{\arabic{footnote}}

\noindent {\bf Introduction:}~ Even though most measurements at the
electroweak scale are consistent with the standard model (SM)
expectations, only a few handful are not quite so, and among them a
notable and long standing one is a more than $3\sigma$ discrepancy in
$a_\mu \equiv (g-2)_\mu / 2$ as $a_\mu^{\rm exp} - a_\mu^{\rm SM} =
(27.05 \pm 7.26) \times 10^{-10}$
\cite{Bennett:2006fi,Roberts:2010cj,Keshavarzi:2018mgv} (see also
\cite{Hagiwara:2011af,Davier:2010nc,Davier:2016iru}). Can
supersymmetry (SUSY) explain this anomaly
\cite{Lopez:1993vi,Chattopadhyay:1995ae,Moroi:1995yh,Carena:1996qa}?
In a general SUSY framework, there are two initial hurdles. First, one
needs rather light (few hundred GeV) sleptons and weak gauginos to
explain this discrepancy, but their lightness may lead to unacceptably
large flavor changing neutral currents (FCNC). The second issue is
which one of the two kinds of loops, bino-slepton or
chargino-sneutrino induced, that generate $a_\mu$ is relatively more
important.\footnote{For a particularly light slepton and bino,
  bino-higgsino-slepton loop can be important. However, in this case,
  the stau tends to be much lighter than the selectron or smuon, and
  it becomes too light to escape collider constraints.}  The dominance
of bino-slepton loops results in large left-right stau mixing which
may lead to charge breaking minimum of the scalar
potential~\cite{Hisano:2010re}. To avoid this problem, we may switch
to the dominance of chargino-sneutrino loops, the necessary condition
for which is a large wino-higgsino mixing. In this case, the lightest
supersymmetric particle (LSP), which is a neutralino in the gravity
mediated scenario, has a strong higgsino admixture, and such a DM
candidate is strongly disfavored by the direct search experiments
\cite{Akerib:2016vxi,Aprile:2017iyp,Cui:2017nnn}.  Also, if the
chargino is heavier than sleptons, the LHC constraints on the chargino
mass is too strong for the chargino induced loop to have any numerical
impact on $a_\mu$ \cite{Hagiwara:2017lse}.  Gauge mediated
supersymmetry breaking (GMSB) models
\cite{Dine:1993yw,Dine:1994vc,Dine:1995ag} have got a distinct
advantage for addressing all the issues. First, gauge interactions
always keep FCNC under control. Second, constraints from DM direct
search experiments cease to apply on wino-higgsino mixing, as in GMSB
the gravitino constitutes the LSP. Third, minimal GMSB with a pair of
messengers transforming in $\five$ and $\fivebar$ representations of
$SU(5)$ yields larger soft masses for sleptons than for the wino, with
grand unified theory (GUT) breaking masses for the messengers.
Thus GMSB naturally
provides an advantageous platform for addressing the $a_\mu$ crisis.

\noindent
Now we face the next level of hurdles within the GMSB framework.  LHC
data tell us that a wino has to weigh above 1.1
TeV~\cite{Aaboud:2018jiw} if it is heavier than sleptons. On the
contrary, if sleptons are heavier than the wino, which is indeed the
case in GMSB with a pair of $\five + \fivebar$ messengers (minimal
GMSB), the wino mass limit weakens to much lower values. Now we recall
that in a generic GMSB framework, to match a heavy stop in the range
$\sim10$\,TeV (necessary to reproduce the observed Higgs boson mass
$m_h \simeq 125$ GeV~\cite{Okada:1990vk, Ellis:1990nz, Haber:1990aw,
  Okada:1990gg, Ellis:1991zd}), the higgsino mass parameter $\mu$ must
be $(3\,\mathchar`-\,4)$ TeV for the correct realization of the weak
scale $M_Z (\simeq 91~{\rm GeV})$, even if the messenger scale is
quite low.  For $\mu$ so large, the chargino is practically a wino,
which in minimal GMSB constitutes the next-to-LSP (NLSP).  The
(weakened) limit on its mass is 460 GeV~\cite{Aaboud:2017mpt}, if
gravitino is in the keV range, arising from non-observation of
disappearing tracks at the LHC.\footnote{The above limit strengthens
  to around 600 GeV for superlight gravitino in the eV
  range~\cite{Aaboud:2018jiw}.}  With such a large $\mu$-parameter,
the higgsino admixture to wino is negligible, and contributions from
the bino-slepton induced loops dominate the SUSY contribution to
$a_\mu$.  However, in this case, it is impossible to explain the
$a_\mu$ discrepancy at a satisfactory level since the sleptons are not
enough light because of the wino mass limit of
460\,GeV~\cite{Yanagida:2017dao}. Also, we face the problem of charge
breaking minimum in the stau-Higgs potential imposing tight
constraints on the parameter space.  The situation does not improve
even in cases with $\ten+\overline{\ten}$ messengers or $\24$
messenger~\cite{Yanagida:2017dao}.  Also, the GMSB model with $SU(3)$
octet and $SU(2)$ triplet
messengers~\cite{Bhattacharyya:2013xba,Bhattacharyya:2013xma} cannot
avoid the current LHC constraints in the region where the $a_\mu$
discrepancy is explained.

\noindent
In this paper, our basic framework is minimal GMSB (with a pair of
$\five + \fivebar$ messengers). However, to circumvent the above
problems, we arrange for a smaller $\mu$ through the introduction of a
new soft SUSY breaking mass (3-4 TeV) of the Higgs bosons.  This
  single parameter ($\delta m_H^2 \equiv \delta m_{H_u}^2 = \delta
  m_{H_d}^2$) brings in a clear advantage to $a_\mu$ as shown in
  Ref.~\cite{Yanagida:2017dao}.  With a small $\mu$, the
wino-higgsino mixing becomes sizable, and chargino-sneutrino loops
dominate the SUSY contribution to $a_\mu$.  Consequently, even with
relatively heavier sleptons one can explain the $a_\mu$
discrepancy. One more difficulty, as a consequence of the most
  updated LHC data, needs to be attended to, which is not considered in
  Ref.~\cite{Yanagida:2017dao}.  With just $\five + \fivebar$
messengers, $B_\mu$ vanishes at the messenger scale. Then $\tan\beta$
is predicted to be rather large ($\sim 50$) leading to rather small
masses of the heavy neutral Higgs bosons (CP-even $H$ and CP-odd $A$)
disfavored by the latest LHC data \cite{Aaboud:2017sjh}. To solve this
last problem we introduce a soft $B_\mu$ parameter (which couples the
two Higgs doublets $H_u$ and $H_d$ in the scalar potential) of size
$\sim 10^5$\,GeV$^2$ at the messenger scale. We demonstrate that the
dynamical origin of the new soft terms, $\delta m_H^2$ and $B_\mu$,
can be traced to a simple extension of the minimal GMSB
superpotential.

\noindent {\bf Our model:}~ In our setup, we must keep provision for GUT breaking effects to create sizable mass
splittings between colored and uncolored SUSY particles, without which
it is impossible to simultaneously satisfy $a_\mu$ and $m_h \simeq
125$ GeV. Now we write the superpotential with $\five + \fivebar$
messenger multiplets as
\begin{eqnarray}
W = (M_L +  k_L Z ) \Psi_L \Psi_{\bar L} +(M_L +  k_D Z ) \Psi_D \Psi_{\bar D}, 
\end{eqnarray}
where $\Psi_L$ and $\Psi_D$ are the $SU(2)$ doublet and $SU(3)$
triplet messengers, respectively, and $Z$ is a SUSY breaking
superfield. It is assumed $\left<Z\right>\ll M_{L,D}$. 
The masses of the colored (uncolored) superpartners are
essentially determined by $\Lambda_D \equiv k_D F_Z/M_D$ ($\Lambda_L
\equiv k_L F_Z/M_L$) as 
\begin{eqnarray}
M_{\tilde b} \simeq \frac{g_1^2}{16\pi^2} \left(\frac{3}{5} \Lambda_L
+ \frac{2}{5} \Lambda_D \right), \ \ M_{\tilde w} \simeq
\frac{g_2^2}{16\pi^2} \Lambda_L, \ \ M_{\tilde g} \simeq
\frac{g_3^2}{16\pi^2} \Lambda_D, \nonumber \\
\tilde m^2 \simeq \frac{2}{(16\pi^2)^2} \left[ {C_2(r_3)} g_3^4
  \Lambda_D^2 + C_2(r_2) g_2^4 \Lambda_L^2 + \frac{3}{5}Q_Y^2 g_1^4
  \Lambda_Y^2 \right], \label{eq:mass_formula}
\end{eqnarray}
where $M_{\tilde b}$, $M_{\tilde w}$ and $M_{\tilde g}$ are the bino,
wino and gluino masses, respectively; $\tilde m^2$ represents soft SUSY
breaking mass-square of a squark, slepton or Higgs doublet; $g_1$,
$g_2$ and $g_3$ are the gauge coupling constants of $U(1)_Y$,
$SU(2)_L$ and $SU(3)_C$; $C_2(r_a)$ is the quadratic Casimir invariant
of the representation $r_a$; $Q_Y$ stands for hypercharge; $\Lambda_Y^2 =
(3/5) \Lambda_L^2 + (2/5) \Lambda_D^2$.
The equality $\Lambda_L=\Lambda_D$ signifies GUT
preserving condition, but it is only by taking $\Lambda_D \gg
\Lambda_L$ one can generate much larger masses for squarks and gluino
than those for sleptons and wino~\cite{Ibe:2012qu}. 
Sleptons, which receive mass from $\Lambda_L$ and $\Lambda_D$, are
naturally heavier than wino, which receives mass from $\Lambda_L$ only.
Satisfaction of $m_h \simeq 125$ GeV
requires the stop (squark) mass to be $\sim$10\,TeV, and as a
corollary, the electroweak symmetry breaking conditions dictate that
the higgsino mass parameter $\mu$ should also be large as
$\sim$3-4\,TeV.  Consequently, the chargino induced loop contribution
to $a_\mu$ is suppressed.  In such a situation, especially light
sleptons, bino and wino are necessary to satisfy $a_\mu$.  However, as
shown in Ref.~\cite{Yanagida:2017dao}, the required spectrum is
excluded by the wino mass limit $(\gtrsim460\,{\rm GeV})$ at the
LHC~\cite{Aaboud:2017mpt} and the vacuum stability constraint on the
Higgs-stau potential~\cite{Hisano:2010re}.
%

\noindent
As a remedy to the above problem, we must arrange for a small $\mu$,
which would enhance the chargino contribution to $a_\mu$. This is
achieved by the introduction of $\delta m_H^2$ at the messenger
scale. But this is not enough for escaping all the LHC constraints.
If the soft parameter $B_\mu = 0$ at the messenger scale, $\tan\beta$
is predicted to be too large $(\gtrsim 50)$. This in turn leads to
rather small masses of the heavy neutral Higgs boson ($\lesssim
1.5\,{\rm TeV}$)~\cite{Yanagida:2017dao}, which is excluded at the LHC
from searches in the $H/A\to\tau\tau$ channel~\cite{Aaboud:2017sjh}.
Therefore, we introduce a non-vanishing $B_\mu$ at the messenger
scale, which enables us to choose smaller $\tan\beta$ to evade the
above constraint. Incidentally, the bino-higgsino-smuon loop
  contribution to $a_\mu$ is numerically not significant in our
  scenario, since the bino is rather heavy due to the contribution
  from $\Lambda_D (\gg \Lambda_L)$ -- see Eq.~(\ref{eq:mass_formula}).

\noindent {\bf Parameters:}~ We deal with five parameters: $(\delta
m_{H}^2, \tan\beta, \Lambda_L, \Lambda_D, M_L=M_D)$. Here, $\delta
m_{H}^2=\delta m_{H_u}^2=\delta m_{H_d}^2$, and we choose $\tan\beta$
to be a free parameter in lieu of a non-vanishing $B_\mu$. A toy
scenario, as an existence proof for $\delta m_H^2$ and $B_\mu$, is
presented later.

\begin{figure}[t]
 \begin{center}
   \includegraphics[width=70mm]{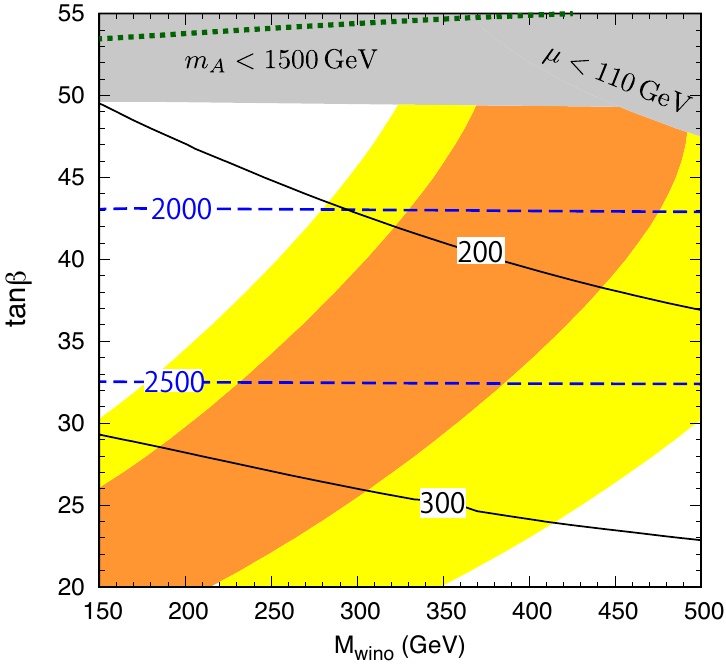}
   \includegraphics[width=70mm]{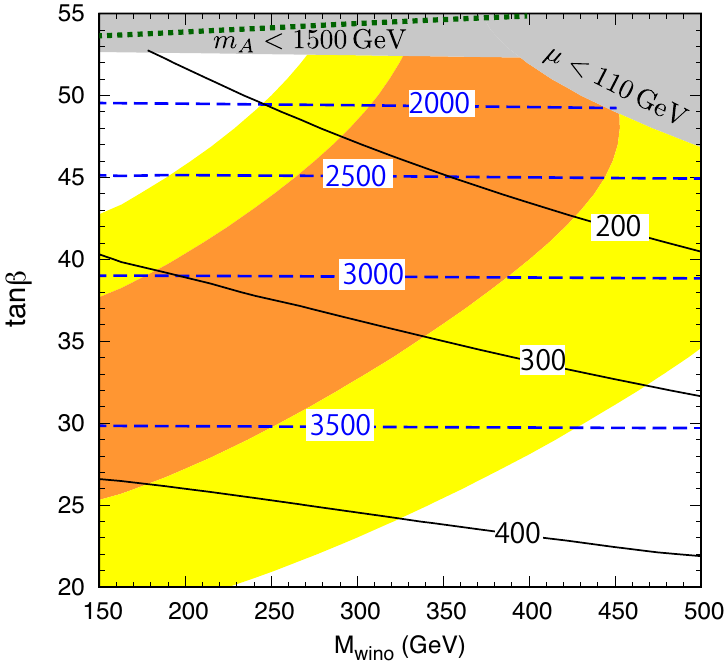}
 \end{center}
  \caption{\em \small The contours of $\Delta a_\mu$, $m_A$ (blue dashed)
    and $\mu$ (black solid), where $m_A$ and $\mu$ are shown in units
    of GeV. In the orange (yellow) regions, the muon $g-2$ is
    explained at $1\sigma\, (2\sigma)$ level.  In the gray regions,
    $\mu<110\,{GeV}$ or $m_A < 1500$\,GeV. On the green dotted
    lines, $B_\mu=0$ at the messenger scale $M_L=M_D$.  In the left
    (right) panel, we take $\Lambda_D=700\,TeV$,
    $M_{L}=M_D=1000$\,TeV and $\delta m_H^2=8.96 \cdot 10^6$\,GeV$^2$
    ($\Lambda_D=1000\,{TeV}$, $M_{L}=M_D=1200$\,TeV and $\delta
    m_H^2=1.64 \cdot 10^7$\,GeV$^2$). Here, $\alpha_s (m_Z) =0.1185$
    and $m_t(pole)=173.34$\,GeV. }
 \label{fig:gm2}
\end{figure}

\begin{table*}[!t]
\caption{\em \small Mass spectra in sample points. We take
  $M_L=M_D$. Although the charginos and neutralinos are quite light,
  they satisfy the LHC constraints~\cite{Aaboud:2018jiw}.  }
\label{tab:sample}
\begin{center}
\begin{tabular}{|c||c|c|c|c|}
\hline
Parameters & Point {\bf I} & Point {\bf II}  & Point {\bf III}  \\
\hline
$M_{D} $ (TeV) & 1000  & 1200  & $1200$  \\
$\Lambda_{D} $ (TeV) & 700  & 1000  & 1000  \\
$\Lambda_L/\Lambda_D$  & $0.18$  & $0.14$  & $0.09$ \\
$\delta m_{H}^2 $ ($10^7$\,GeV$^2$) & 0.89  & 1.64  & 1.66  \\
$\tan\beta$ & 40 & 40 & 25 \\
\hline
%
Particles & Mass (GeV) & Mass (GeV)& Mass (GeV)  \\
\hline
$\tilde{g}$ & 5150 & 7510 & 7510   \\
$\tilde{q}$ & 6500 & 9080 & 9080 \\
$\tilde{t}$ & 6020 & 8430 & 8460 \\
$\tilde{\chi}_{1}^\pm$ & 266 & 243  & 178 \\
$\tilde{\chi}_{2}^\pm$ & 385 & 401  & 294 \\
$\tilde \chi_1^0$         & 261 & 238  & 174 \\
$\tilde{\chi}_2^0$        & 310 & 268 & 233  \\
$\tilde \chi_3^0$         & 380 & 399  & 290\\
$\tilde{\chi}_4^0$       & 534 & 765 & 723  \\
$\tilde{e}_{L, R}$       & 556,\,796 & 689,\,1120  & 575,\,1110 \\
$\tilde{\tau}_{1,2}$     & 449,\,650&  535,\,935  & 501,\,1030 \\
$H/A$ & 2180 & 2920 & 3690 \\
$h_{\rm SM\mathchar`-like}$ & 125.9 &  127.5  & 127.8 \\
\hline
$\mu$ (GeV)  & 296  & 254  & 218 \\
$B_\mu (M_D)$ (10$^5$\,GeV$^2$)  & 1.07  & 2.03  & 5.55 \\
$\Delta a_\mu (10^{-10})$  & 29.9  & 23.1  & 22.2 \\
\hline
\end{tabular}
\end{center}
\end{table*}

\noindent
In Fig.~\ref{fig:gm2}, we show the contours of $\Delta a_\mu$, $m_A$
and $\mu$, on the $M_{\tilde w}$-$\tan\beta$ plane, where $M_{\tilde
  w}$ $(\mu)$ is the wino mass (higgsino mass) at the stop mass scale,
$m_A$ is the CP-odd Higgs boson mass, and $\Delta a_\mu$ is the SUSY
contribution to the muon $g-2$. We have calculated $\Delta a_\mu$ and
the mass of the SM-like Higgs boson using FeynHiggs
2.14.1~\cite{Heinemeyer:1998yj,Heinemeyer:1998np,Degrassi:2002fi,Frank:2006yh,Hahn:2013ria}. The
SUSY mass spectra are evaluated using SOFTSUSY
3.7.4~\cite{Allanach:2001kg} with appropriate modifications.  In the
orange (yellow) regions, $a_\mu$ is explained at 1$\sigma$ (2$\sigma$)
level. The higgsino mass parameter $\mu$ lies within the range of
200-400 GeV.
In the viable regions for $a_\mu$ explanation, $B_\mu$ is
$\mathcal{O}(10^5)$\,GeV$^2$,
$\Lambda_L/\Lambda_D=\mathcal{O}(0.1)$, and $m_h \approx$126\,GeV
(127\,GeV) in the left (right) panel.  The required GUT breaking
effect ($\Lambda_L \neq \Lambda_D$) is, notably, rather mild.

\noindent
In Table.~\ref{tab:sample}, the mass spectra in sample points are
shown. All the points are consistent with the muon $g-2$ result at
1$\sigma$\,level. The heavy Higgs boson masses calculated by FeynHiggs
are sufficiently large on all the points.  The mass spectra of the
Heavy neutral Higgs boson are large enough to avoid the LHC constraint
$(m_A \gtrsim 1.5\,{\rm TeV}$ for $\tan\beta \gtrsim
40)$~\cite{Aaboud:2017sjh}. The wino-like chargino and neutralino,
$\chi_2^\pm$ and $\chi_3^0$, decay into $\chi_1^0$, $\chi_2^0$ and
$\chi_1^\pm$ via emitting $W/Z$ boson, and even when they are as light
as $\sim200$\,GeV the LHC constraints are
satisfied~\cite{Aaboud:2018jiw}. Note that we have focused on the cases
  where the messenger scale is as low as $\sim10^6$\,GeV.  This is
  because for a large messenger scale, e.g. $\sim10^{11}$\,GeV, the
  stau becomes lighter eventually turning into NLSP due to radiative corrections in the
  region consistent with the muon $g-2$. This light stau is excluded
  by the LHC
  searches~\cite{ATLAS:2014fka,Khachatryan:2016sfv}.\footnote{
For a moderately large messenger scale, e.g. $10^8$\,GeV, we still
have a region consistent with the muon $g-2$. However, the lightness
of the stau renders this region to be extremely narrow.}

It is also important to notice that we have only considered the
  cases where the NLSP is stable inside the LHC detector.  This
  follows from the observation that the gravitino weighing less than
  $\sim 10$\,keV is strongly constrained by the Lyman-$\alpha$ forest
  data~\cite{Baur:2015jsy}. For a heavier gravitino the NLSP decay
  length turns out to be longer than the detector size.  To see this,
  we first estimate the gravitino mass as
\begin{eqnarray}
m_{3/2} \simeq \frac{F_Z}{\sqrt{3} M_P} \approx 17\,{\rm keV}
\left(\frac{\Lambda_D}{700\,{\rm TeV}} \right)
\left(\frac{M_D}{1000\,{\rm TeV}}\right)
\left(\frac{k_D}{10^{-2}}\right)^{-1},
\end{eqnarray}
where $M_P \approx 2.4 \times 10^{18}\,{\rm GeV}$ is the reduced
Planck mass.\footnote{Although the gravitino lighter than 4.7\,eV
 suffers no cosmological constraints~\cite{Osato:2016ixc}, such a
  light gravitino requires $k_D$ to be much larger than $4\pi$ to
  compensate for the smallness of $F_Z$ for reproducing the allowed
  superpartner spectra. } With the
gravitino mass of $\mathcal{O}(10)$\,keV, the decay length of the
higgsino-like NLSP is given by
\begin{eqnarray}
c \tau \approx 140\, {\rm m} \times \left(\frac{\mu}{250\,{\rm
    GeV}}\right)^{-5} \left(\frac{m_{3/2}}{20\,{\rm keV}}\right)^2,
\end{eqnarray}
which is too large to be constrained at the
LHC~\cite{Liu:2015bma,Calibbi:2018fqf}.
Therefore, the higgsino-like NLSP can be regarded effectively as a stable
particle in the collider time scale. This argument also applies to
the wino-like NLSP case as long as the gravitino mass is heavier than
$\mathcal{O}(10)$\,keV.
 
\noindent {\bf An ultraviolet completion:}~ Here we show an example
model for generating the new soft parameters, namely, $\delta m_H^2$
and $B_\mu$, in the Higgs sector.  The relevant superpotential is
\begin{eqnarray} 
W = \frac{\kappa}{2} Z X^2 + M_{X} X \overline{X} + 
\lambda X H_u H_d \, .
\end{eqnarray}
It generates,  
as in \cite{Ibe:2012qu},
$\delta m_{H}^2 \equiv \delta m_{H_u}^2=\delta m_{H_d}^2$, where
\begin{eqnarray}
\delta m_H^2 \simeq \frac{|\lambda|^2 |\kappa|^2}{32\pi^2}
\frac{|F_Z|^2}{M_{X}^2}, \label{eq:mhsq}
\end{eqnarray}
which is required to be $\mathcal{O}(10^7)$\,GeV$^2$.  Note that $W$
has a $R$ symmetry which prevents the generation of the soft $B_\mu$
term and the trilinear $A$ parameter, as long as $\langle Z \rangle =
0$. 
However, it may be possible that the $\left<Z\right> \neq 0$ due to unknown hidden sector dynamics,and the $R$ symmetry is spontaneously broken. Otherwise, one can introduce a term $m_{X}X^2/2$ in W, where $m_{X}$ may be considered to be small as this term is an explicit $R$-violating one. The induced $B_\mu$ and trilinear $A$-terms are given by~\footnote{In order to generate $B_\mu \gtrsim 10^5$\,GeV$^2$, a strongly interacting theory with $\lambda \gg 1$ needs to be considered.}
\begin{eqnarray}
B_\mu/\mu, A_{u,d} \simeq \frac{|\lambda|^2 \kappa}{32\pi^2}
\frac{F_Z}{M_X} \frac{m_X}{M_X} \left[1  + \mathcal{O}\left(\frac{m_X^2}{M_X^2}\right) \right].
\end{eqnarray}

\noindent
On the other hand, if $\langle Z \rangle = 0$, one may add singlet superfields $S$ and $\overline{S}$
 to write a new superpotential
\begin{eqnarray} 
W' = \lambda_u \Psi_{L}
  H_u S + \lambda_d \Psi_{\bar L} H_d \overline{S} + M_S S
  \overline{S} \, .
\end{eqnarray}
The superpotential $W'$ generates the $B_\mu$ term
as~\cite{Asano:2016oik}
\begin{eqnarray}
B_\mu \simeq  \frac{\lambda_u \lambda_d}{16\pi^2} \Lambda_L^2 R(x),
\end{eqnarray}
where $x=M_S/M_L$ and $R(x)$ is a loop function with
$R(1)=1/3$. 
Note that $W'$ also generates $m_{H_u}^2$ and $m_{H_d}^2$. 
But their magnitudes are too small, compared to what we get from Eq.\,(\ref{eq:mhsq}) induced by the superpotential $W$, 
when we satisfy the phenomenological requirement $B_\mu=\mathcal{O}(10^5)$\,GeV$^2$ to be able to take $\tan\beta \sim 40$. Furthermore, the $W'$ induced $\mu$ and $A$-terms are negligible.  

\noindent {\bf Conclusion:}~ As the explanation of $a_\mu$ requires
light sleptons and weak gauginos, gauge mediation provides an
attractive framework by keeping FCNC under control, creating a slepton
/ weak gaugino mass hierarchy that is less constrained at the LHC, and
separating the weak gaugino sector from the dark matter (gravitino)
search.  We have shown that a simple extension of minimal gauge
mediation by introducing soft SUSY breaking parameters for the Higgs
sector can achieve the non-trivial goal of explaining the $a_\mu$
discrepancy. We also demonstrated that the newly introduced parameters
can be justified in an ultraviolet complete theory.
We claim that our scenario stands out at least as the only GMSB
model, to the best of our knowledge, that can explain $a_\mu$ and
simultaneously satisfy all the other constraints. A low $\mu$ (few
hundred GeV) that arises in our scenario also provides an impetus for
a dedicated higgsino search at a future linear collider. 

\noindent {\bf Acknowledgements:}~ G.B. thanks Kavli IPMU for
hospitality when the work was done.  G.B. acknowledges support of the
J.C.~Bose National Fellowship from the Department of Science and
Technology, Government of India (SERB Grant No.\ SB/S2/JCB-062/2016).
This work is supported by JSPS KAKENHI Grant Numbers 
JP26104001 (T.T.Y), JP26104009 (T.T.Y), JP16H02176 (T.T.Y),
JP17H02878 (T.T.Y), JP15H05889 (N.Y.),
JP15K21733 (N.Y.), JP17H05396 (N.Y.), JP17H02875 (N.Y.), and by World
Premier International Research Center Initiative (WPI Initiative),
MEXT, Japan (T.T.Y.).


\end{document}